\documentclass[twocolumn,secnumarabic,amssymb, nobibnotes, aps, prd]{revtex4}
\usepackage{graphicx}

\begin{document}
\title{The fundamental obscurity in quantum mechanics. \\ Why it is needed to shout "wake up"}
\author{V.V. Aristov and  A.V. Nikulov}
\affiliation{Institute of Microelectronics Technology and High Purity Materials, Russian Academy of Sciences, 142432 Chernogolovka, Moscow District, RUSSIA.} 
\begin{abstract} The orthodox quantum mechanics can describe results of observation but not an objective reality. Misunderstanding of its positivistic essence results to numerous mistaken publications, in particular about a possibility of a real quantum computer. 
 \end{abstract}

\maketitle

\narrowtext

\section*{Introduction}

The inevitable transition from classical world to quantum world because of the progressive miniaturization of nanostructures has prompted such fundamentally new idea as quantum computing \cite{Aristov1}. The widespread interest in this idea seems quite valid. The minimal sizes of nanostructures come nearer to atomic level and subsequent miniaturization will not be possible in the near future. Therefore exponential increase of calculating resources with number of quantum bits has provoked almost boundless enthusiasm. This exponential increase seems possible thanks to the principle of superposition of states, interpreted \cite{Aristov2} as the cardinal positive principle of quantum mechanics. Feynman \cite{Aristov3} proposed universal simulation, i.e. a purpose-built quantum system which could simulate the physical behaviour of any other, noting that the calculation complexity of quantum system increases exponentially with number $n$ of its elements. A system of $n$ elements $\psi_{i}$ with two observable values $+1$ or $-1$ should describe by the superposition
$$\psi = \gamma_{1} \psi_{1}(+)\psi_{2}(+) \cdot \cdot \cdot  \psi_{i}(+) \cdot \cdot \cdot  \psi_{n}(+) + \cdot \cdot \cdot  $$  
$$+ \gamma_{2} \psi_{1}(-)\psi_{2}(+) \cdot \cdot \cdot    \psi_{i}(+) \cdot \cdot \cdot \psi_{n}(+)  + \cdot \cdot \cdot  $$ 
$$+ \gamma_{j} \psi_{1}(-)\psi_{2}(-) \cdot \cdot \cdot   \psi_{i}(-) \cdot \cdot \cdot   \psi_{n}(+) +  \cdot \cdot \cdot  $$
$$+ \gamma_{N} \psi_{1}(-)\psi_{2}(-) \cdot \cdot \cdot \psi_{i}(-) \cdot \cdot \cdot \psi_{n}(-) \eqno{(1)}$$ 
with $N - 1 = 2^{n} - 1$ independent variables $\gamma_{j}$. Feynman gave arguments which suggested that quantum evolution of (1) could be used to compute certain problems more efficiently than any classical computer. Indeed, it seems possible to change $N = 2^{n}$ variables $\gamma_{j}$ with help of an influence on a single element $\psi_{i}$ of the quantum system (1). This feature is referred to as quantum parallelism and represents a huge parallelism because of the exponential dependence on n: only 1000 elements, quantum bits, seems to replace $2^{1000} = 10^{301}$ classical processors \cite{Aristov1}. 

The universal simulation proposed by Feynman could not result in the idea of universal quantum computer, possibility of which was substantiated by Deutsch \cite{Aristov4}. In the early 1990's several authors sought computational tasks which could be solved by a quantum computer more efficiently than any classical computer. Shor has described in 1994 \cite{Aristov4} an algorithm which was not only efficient on a quantum computer, but also addressed a central problem in computer science: that of factorising large integers. Armed with Shor's algorithm, it now appears that such a fundamental significance is established, by the following argument: either nature does allow a device to be run with sufficient precision to perform Shor's algorithm for large integers (greater than, say, a googol, $10^{100}$), or there are fundamental natural limits to precision in real systems. Both eventualities represent an important insight into the laws of nature.

\section {Could quantum computer be possible as a real device?}
The Shor's algorithm has provoked numerous publications \cite{Aristov6,Aristov7,Aristov8} growing rapidly about quantum computing and colossal efforts applied to creation of a real quantum computer. Because of this enthusiasm only few experts venture to doubt in the reality of quantum computer. Nevertheless this doubt is very valid. There is important to note first of all that the substantiation by Deutsch the possibility of a real universal quantum computer must be connected with his belief in the "Many Universes Theory" of quantum physics \cite{Aristov9}. According to this idea when a particle changes, it changes into all possible forms, across multiple universes. Deutsch proving the connection between the reality of the quantum computer and the existence of the parallel universes asks in his book \cite{Aristov10} {\it those who is still declined to count, that there is only one universe}: {\it "When Shor's algorithm has factorized number, having involved about $10^{500}$ computing resources which can be seen where this number was factorized on  multipliers? About $10^{80}$ atoms exist in whole seen universe, the number is insignificant small in comparison with $10^{500}$"}. 

The unreality of the quantum computer only in one universe can be understood from the fact that the exponential increase of the number $N - 1 = 2^{n} - 1$ of independent variables of the quantum register (1) with the number $n$ of quantum bits can be possible only at the invalidity of the relation 
$$\psi_{i} = \alpha_{i} \psi_{i}(+)  + \beta _{i}\psi_{i}(-); \ \   \alpha_{i}^{2} + \beta _{i}^{2} = 1 \eqno{(2)}$$
for the superposition of states of each quantum bit. The invalidity of (2) can not be real only in one universe logically: the quantum bit (2) has only two states $\psi_{i}(+)$, $\psi_{i}(-)$ and the sum of their probability $\alpha_{i}^{2}$, $\beta _{i}^{2}$ must be equal unity. The relations $\alpha_{i}^{2} + \beta _{i}^{2} = 1$ for each quantum bit reduce the quantum register (1) to the classical one with $N \approx  n $.

\section {Irremediably conflict of the entanglement with realism}
The invalidity of (2) at the entanglement or EPR correlation was revealed by the opponent \cite{Aristov11,Aristov12} of the Copenhagen interpretation in order to prove the incompleteness of the quantum description of physical reality. The EPR paradox has demonstrated irremediably conflict of the superposition principle with local realism: the collapse of the superposition 
$$\psi_{EPR} = \alpha_{EPR} \psi_{1}(+)\psi_{2}(-)  + \beta _{EPR}\psi_{1}(-)\psi_{2}(+) \eqno{(3)}$$
describing the EPR pair at observation of one particle 1 
$$\psi_{EPR} = \psi_{1}(+)\psi_{2}(-)  \eqno{(4)}$$
implies instantaneous change of the state of the other particle $2$, irrespective of the distance between these particles.  Schrodinger described the EPR correlation (3) as {\it entanglement of our knowledge} \cite{Aristov12} because of this conflict. The identification of the state vector (1,2,3) with "knowledge of the system" by Heisenberg is defined in \cite{Aristov13} as the fourth principal element of the Copenhagen interpretation. This information interpretation is enough natural because the collapse from (3) to (4) describes the change of our knowledge at observation. It eliminates simple non-locality problem {\it non-locality of the first kind} according to \cite{Aristov13}, which is at the description. The famous work by John Bell \cite{Aristov14} has allowed to reveal the non-locality on a more deeper level, at observations. Interpretations become irrelevant because real observations are involved in this {\it non-locality of the second kind} \cite{Aristov13}.

The experimental evidences \cite{Aristov15} of violation of the Bell's inequalities testify to the observation of the EPR correlation. But it is mistake to conclude that the EPR correlation can exist really because {\it it is a gross violation of relativistic causality} \cite{Aristov16}. For Bell the violation of the Bell's inequality was {\it "the real problem with quantum theory: the apparently essential conflict between any sharp formulation and fundamental relativity"} \cite{Aristov16}. This troubling conflict between the empirically verified predictions of quantum theory and the notion of local causality that is motivated by relativity theory is discussed by experts \cite{Aristov17,Aristov18, Aristov19,Aristov20}. Some of they hope \cite{Aristov17,Aristov18} that this conflict can be overcome at the cost of renunciation of realism and determinism, whereas the other one \cite{Aristov19,Aristov20} state that even this can not be possible. Mermin confessed in 2001 \cite{Aristov21}: {\it "Until quite recently I was entirely on Bell's side on the matter of knowledge-information. But then I fell into bad company. I started hanging out with the quantum computation crowd, for many of whom quantum mechanics is self-evidently and unproblematically all about information"}. Because of his associations with quantum computer scientists he has {\it "come to feel that "Information about what?" is a fundamentally metaphysical question that ought not to distract tough-minded physicists"} \cite{Aristov21}. Ghirardi, calling in question this point of view, reminds \cite{Aristov22} that Bell {\it "refused to consider such a position unless \cite{Aristov23}, in advance, one would have answered to two basic (for him) questions: whose information?, and: information about what?"}  This controversy by the experts testifies that the EPR correlation, and consequently quantum computer, can not be real in the reality of single universe.  

\section {The principle of superposition and positivism}
Thus, the EPR correlation and violation of the Bell's inequalities \cite{Aristov15} testify rather to the fundamental obscurity in quantum mechanics than to a possibility of a real equipment. Ironically, numerous authors \cite{Aristov1,Aristov6,Aristov7,Aristov8} are sure that this unreal principle introduced by the opponent \cite{Aristov11,Aristov12} of the Copenhagen interpretation can be a basis of a real device, quantum computer. This mass delusion could be possible because of blind admiration on the progress of physics and engineering of the XX century developed thanks to quantum mechanics. But as Bell noted \cite{Aristov16} {\it "This progress is made in spite of the fundamental obscurity in quantum mechanics. Our theorists stride through that obscurity unimpeded... sleepwalking?"} He said \cite{Aristov16}: {\it "The progress so made is immensely impressive. If it is made by sleepwalkers, is it wise to shout 'wake up'? I am not sure that it is. So I speak now in a very low voice"}. The authors \cite{Aristov1,Aristov6,Aristov7,Aristov8} and others do not understand that superposition of state can not be real and that this principle can be valid only in the limits of the positivism point of view, according to which quantum mechanics can describe only phenomena but no a reality. The numerous mistaken publications, possible because of this lack of understanding of the fundamental obscurity connected with the positivistic essence of quantum mechanics, compel to shout "wake up".


\begin{thebibliography}{99}

\bibitem{Aristov1} A. Steane, {\em Rept.Prog.Phys. } {\bf 61}, 117 (1998).

\bibitem{Aristov2} L. D. Landau and E. M. Lifshitz, {\em Quantum Mechanics: Non-Relativistic Theory}. Volume 3, Third Edition, Elsevier Science, Oxford, 1977.

\bibitem{Aristov3} R. P. Feynman, {\em Int. J. Theor. Phys.} {\bf 21} 467 (1982). 

\bibitem{Aristov4} D. Deutsch, {\em Proc. Roy. Soc. Lond. A} {\bf 400}, 97 (1985)

\bibitem{Aristov5} P. W. Shor, in Proc. 35th Annual Symp. on Foundations of Computer Science, Santa Fe, IEEE Computer Society Press (1994); E- print: quantph/ 9508027

\bibitem{Aristov6} M.A. Nielsen, and I.L. Chuang, {\em Quantum Computation and Quantum Information}. Cambridge University Press, 2000.

\bibitem{Aristov7} D. Bouwmeester, A. Ekert, and A. Zeilinger (Eds.), {\em The Physics of Quantum Information. Quantum Cryptography. Quantum Teleportation. Quantum Computation.} Springer, Berlin, Heidelberg, 2000.

\bibitem{Aristov8} T. D. Ladd, F. Jelezko, R. Laflamme, Y. Nakamura, C. Monroe and J. L. O'Brien, {\em Nature} {\bf 464}, 45 (2010).

\bibitem{Aristov9} Q. Norton, The Father of Quantum Computing. http://www.wired.com/science/discoveries/news/2007 /02/72734.

\bibitem{Aristov10} D. Deutsch, {\it The Fabric of Reality}. The Penguin Press, 1997.

\bibitem{Aristov11} A. Einstein, B. Podolsky, and N. Rosen, {\em Phys. Rev.} {\bf 47}, 777 (1935).

\bibitem{Aristov12} E. Schrodinger, {\em Naturwissenschaften} {\bf 23}, 807 (1935); {\em Proc. Cambridge Phil. Soc.} {\bf 31}, 555  (1935).

\bibitem{Aristov13} J.G. Cramer, {\it Rev. Mod. Phys.} {\bf 58}, 647 (1986).

\bibitem{Aristov14} J. S. Bell, {\em Physics} {\bf 1}, 195 (1964).

\bibitem{Aristov15} A. Aspect et al.,  {\em Phys. Rev. Lett.} {\bf 47} 460 (1981);  {\bf 49}, 91 (1982);  {\bf 49}, 1804 (1982).

\bibitem{Aristov16} J. S. Bell, "Speakable and unspeakable in quantum mechanics. Collected papers on quantum philosophy", Cambridge University Press, Cambridge, 1987, p. 169

\bibitem{Aristov17} N.D. Mermin, {\em Am. J. Phys.} {\bf 66(9)}, 753 (1998); arXiv: quant-ph/9801057 

\bibitem{Aristov18} A. Zeilinger, {\em  Nature} {\bf 438}, 743 (2005)

\bibitem{Aristov19} T. Norsen, {\em Found. Phys.}  {\bf 39}, 273 (2009). 

\bibitem{Aristov20} GianCarlo Ghirardi, {\em Found. Phys.} {\bf 40}, 1379 (2010).

\bibitem{Aristov21} N. D. Mermin, in {\em Quantum (Un)speakables: Essays in Commemoration of John S. Bell}. Eds. Reinhold Bertlmann and Anton Zeilinger, Springer Verlag, 2002; arXiv: quant-ph/0107151

\bibitem{Aristov22} Giancarlo Ghirardi, {\it Found. Phys.} {\bf 38}, 1011 (2008).

\bibitem{Aristov23} J.S. Bell, {\em Physics World}, {\bf 3}, 33-40 (1990).

\end{thebibliography}
\end{document}